\documentclass[10pt,letterpaper]{article}
\usepackage{graphics}
\usepackage{amsmath}
\usepackage{subfigure}
\usepackage[pdftex]{graphicx}
\usepackage{color}
\usepackage{hyperref}

\hypersetup{colorlinks,
            linkcolor=black,
            anchorcolor=black,
            citecolor=magenta
            }
\begin{document}

%%%%%%%%%%%%%%%%%% title page information %%%%%%%%%%%%%%%%%%
\begin{center}
{\Large \textbf{ Simultaneous measurement of quality factor and wavelength shift by phase shift microcavity ring down spectroscopy}}\\

{\normalsize M. Imran Cheema$^{1*}$, Simin Mehrabani$^2$, Ahmad A. Hayat$^3$, Yves-Alain Peter$^3$, Andrea M. Armani$^2$ and Andrew G. Kirk$^{1}$}\\

 {\footnotesize $^1$ECE dept., McGill University, 3480 University Street, Montreal, Canada H3A 2A9\\
$^2$Mork Family Department of Chemical Engineering and Materials Science, University of Southern California, Los
Angeles, California 90089, USA\\
$^3$Engineering Physics Dept. Ecole Polytechnique, Montreal, Canada}\\
{\scriptsize $^*$imran.cheema@mail.mcgill.ca} %% email address is required
\end{center}


\section{abstract}
Optical resonant microcavities with ultra high quality factors are widely used for biosensing. Until now, the primary method of detection has been based upon tracking the resonant wavelength shift as a function of biodetection events. One of the sources of noise in all resonant-wavelength shift measurements is the noise due to intensity fluctuations of the laser source. An alternative approach is to track the change in the quality factor of the optical cavity by using phase shift cavity ring down spectroscopy, a technique which is insensitive to the intensity fluctuations of the laser source.  Here, using biotinylated microtoroid resonant cavities, we show simultaneous measurement of the quality factor and the wavelength shift by using phase shift cavity ring down spectroscopy. These measurements were performed for disassociation phase of biotin-streptavidin reaction. We found that the disassociation curves are in good agreement with the previously published results.  Hence, we demonstrate not only the application of phase shift cavity ring down spectroscopy to microcavities in the liquid phase but also simultaneous measurement of the quality factor and the wavelength shift for the microcavity biosensors in the application of kinetics measurements.
\section{Introduction}
The last two decades have seen tremendous progress towards the development of real time, label free and miniature optical biosensors. Researchers have demonstrated numerous techniques based on the surface plasmon resonance, interferometers, waveguides, microcavities, optical fibers, and photonic crystals \cite{Fan_2008}.  Among these different approaches, the large photon lifetime (Quality factor of $10^6-10^9$) of microcavities makes them a strong candidate for ultrasensitive biosensing, as the circulating photons will sample a {biodetection} event many times. {The first microcavity biosensor was demonstrated by Vollmer et. al \cite{Vollmer_2002} in 2002, in which they detected proteins by measuring change of the resonant wavelength of a microsphere as a function of the binding event.} Since then researchers have  transitioned the microcavity-based sensors towards diagnostics platforms and protein behavior detection \cite{Washburn_2011, Soteropulos_2011}.

Until now, the majority of work using microcavity biosensors focused on correlating the change in the resonant wavelength to a {biodetection} event on the surface of the microcavity. However, the binding event will also influence the photon lifetime or the ring down time (and hence quality factor) of the microcavity \cite{Cheema_2011}. Apart from few instances, where quality factors have been determined based upon linewidth measurements \cite{Ilchenko_2002,Armani_2006}, researchers have not sought to measure this property, and  thus important information about the {biodetection} event is lost.  Moreover, in all configurations of the microcavity biosensors used so far, random intensity fluctuations of the laser source add noise into the final measurement and hence the overall performance of the biosensor is degraded \cite{White_2008,Hu_2009}. These issues can be overcome by using the microcavity as the key component in a phase shift cavity ring down spectroscopy (PS-CRDS) measurement.

{Cavity ring down spectroscopy (CRDS) was first demonstrated in 1984 by Anderson et. al \cite{Anderson_1984} in which laser light was injected into a free space optical cavity and after reaching a predefined transmission level at the cavity output, the laser was switched off by using a Pockels cell. The decay rate of the output light was then correlated with reflectivity of the mirrors.} In 1988, O'Keefe et. al \cite{Keefe_1988} transformed CRDS into a sensing method by measuring the absorption of molecular oxygen filled in a free space optical cavity.  Since that time, many variants of the original CRDS technique have been demonstrated, and now CRDS is a well established technique for absorption measurements in gaseous phase \cite{Berden_2009}.  One of the major advantages of this technique is its insensitivity to the intensity fluctuations of the laser source, since the decay rate is measured instead of the absolute intensity of the ring down time signal \cite{Berden_2009}. However, in order to extract the decay rate, fitting algorithms are applied \cite{Lehmann_2009} to the ring down time signal, and as a result, a noise term is added to the final result.  To overcome the fitting noise whilst remaining insensitive to the intensity fluctuations of the laser source, PS-CRDS can be used. This technique was first developed by Herbelin et.al \cite{Herbelin_1980} for measuring reflectance of mirrors and then later Englen et. al \cite{Engeln_1996} successfully applied it to investigate the absorption of vibration states of molecular oxygen. In PS-CRDS, intensity modulated light is injected into the cavity which undergoes a phase change and decrease in modulation depth as it comes out of the cavity. Both of these parameters are related to the ring down time of the cavity.

With the success of CRDS in gaseous phase, the technique is also emerging as a new way for performing absorption measurements in the liquid phase \cite{Sneppen_2008} and for biosensing \cite{Tarsa_2008}. The first application of CRDS biosensing involved a resonator made of an optical fiber with couplers at its ends\cite{Tarsa_2004}. In this work, the researchers used taper section of the fiber as a sensing medium where the evanescent tail of the cavity modes interacted with the analyte molecules.  An alternative, inexpensive design is the use of fiber in a loop with a gap between the two ends for the insertion of a micro-fluidic chip.  Loock et. al. applied PS-CRDS to the fiber loops to detect $\mu M$ concentrations of chemicals in liquid samples \cite{Loock_2004}. However, fiber loop cavities have significantly lower quality factors than silica spheres or toroidal microcavities, resulting in significantly shorter photon lifetimes.  Thus, by combining the advantages of PS-CRDS with an ultra-high-Q microcavity transducer, this sensing modality will be greatly improved. {With this motivation, Barnes et.al \cite{Barnes_2008} applied PS-CRDS to silica microspheres for determining their losses in air. To determine the ring down times, they did linear curve fitting of phase change as a function of modulation frequency. Such a measurement requires the use of a small angle approximation of Eq. (\ref{Eqn:PS_CRDS}) (see section 2.1) and will be marginally correct for only small phase changes. The experimental scheme that was employed is not well-suited for biosensing application as it is not possible for that setup to track the quality factor or the resonant wavelength, as a function of a {biodetection} event, at multiple modulation frequencies.} Therefore, the goal of our research is to show, 1) the application of PS-CRDS to microcavities in the liquid phase, and 2) to simultaneously track changes in the quality factor (ring down time) and the resonant wavelength shift as function of a {biodetection} event by PS-CRDS in a microcavity biosensor.

In our work, we used a bioconjugated \cite{Hunt_2010} microtoroid optical resonator \cite{Armani_2003} to specifically detect unbinding of streptavidin (Fig. \ref{fig:setup}). We tracked both change in the Q and the resonant wavelength by using PS-CRDS and we found that the disassociation curves of biotin-streptavidin are in good agreement with the previously published results.

\section{Phase shift cavity ring down spectroscopy (PS-CRDS)}
\subsection{Theory}

Following the similar procedure as outlined in \cite{Berden_2009, Engeln_1996}  for free space optical cavities, PS-CRDS equations for a microcavity coupled with a tapered optical fiber can be developed.  The intensity modulated light entering the fiber can be represented by :
\begin{equation}
I_{input}=I_o(1+\alpha_{input} \sin \omega t)
\label{Eqn:I_input}
\end{equation}
where $\alpha$ is the modulation depth and $\omega$ is the modulation frequency.\\
Depending upon the coupling regime \cite{Spillane_2004}, a fraction of light will be coupled to the microcavity. If $I_1$ represents the uncoupled light and $I_2$ represents the coupled light, then the modulated light entering into the microcavity can be written as:
\begin{equation}
I_{cavity}=I_2(1+\alpha_{input} \sin \omega t)
\label{Eqn:I_c}
\end{equation}
If $\tau$ represents the ring down time of the microcavity, then its impulse response will be:
\begin{equation}
h(t)=C \text{ } exp\left(-\dfrac{t}{\tau}\right)
\label{Eqn:IR}
\end{equation}
Therefore intensity at the fiber output will be
\begin{eqnarray}
  I_{output}(t)&=&I_1(1+\alpha_{input} \sin \omega t)+I_{cavity} * h(t)\\
            & =& I_1(1+\alpha_{input} \sin \omega t) + C\int_{-\infty}^t I_2(1+\alpha_{input} \sin \omega \acute{t})exp\left(-\dfrac{t-\acute{t}}{\tau}\right)d\acute{t}
\label{Eqn:I_output}
\end{eqnarray}
where $\acute{t}$ is dummy variable for performing convolution. Under steady state conditions \cite{Spillane_2004} for a microcavity coupled with a tapered fiber, the value of C can be found by applying the principle of conservation of energy, i.e. total {intensity} coupled to the cavity in a time interval must be equal to the {intensity} coupled out in that time interval i.e.
\begin{eqnarray}
C\int_{-\infty}^t I_2exp\left(-\dfrac{t-\acute{t}}{\tau}\right)d\acute{t}&=& I_2\\
\Rightarrow  C &=& \dfrac{1}{\tau}
\label{Eqn:C}
\end{eqnarray}
Therefore,
\begin{equation}
I_{output}(t)=I_1(1+\alpha_{input} \sin \omega t)+I_2(1+\alpha_{output} \sin (\omega t + \phi))
\label{Eqn:I_o_final}
\end{equation}
where,
\begin{equation}
	\tan\phi=-\omega \tau
	\label{Eqn:PS_CRDS}
\end{equation}
\begin{equation}
	\alpha_{output}=\dfrac{\alpha_{input}}{\sqrt{1+ \omega^2 \tau^2}}
	\label{Eqn:MD_PS_CRDS}
\end{equation}
Hence, the ring down time $(\tau)$ of the microcavity can be found by taking the ratio of in phase and out of phase sinusoidal components at the fiber output. The quality factor then can easily be extracted by the following relation:
\begin{equation}
     Q = \dfrac {2 \pi c \tau}{\lambda_{resonant}}
	\label{Eqn:Q}
\end{equation}
where $\lambda_{resonant}$ is the resonant wavelength of the microcavity.
\subsection{Experimental Setup}
\begin{figure}[htbp]
\centering\includegraphics[width=13cm]{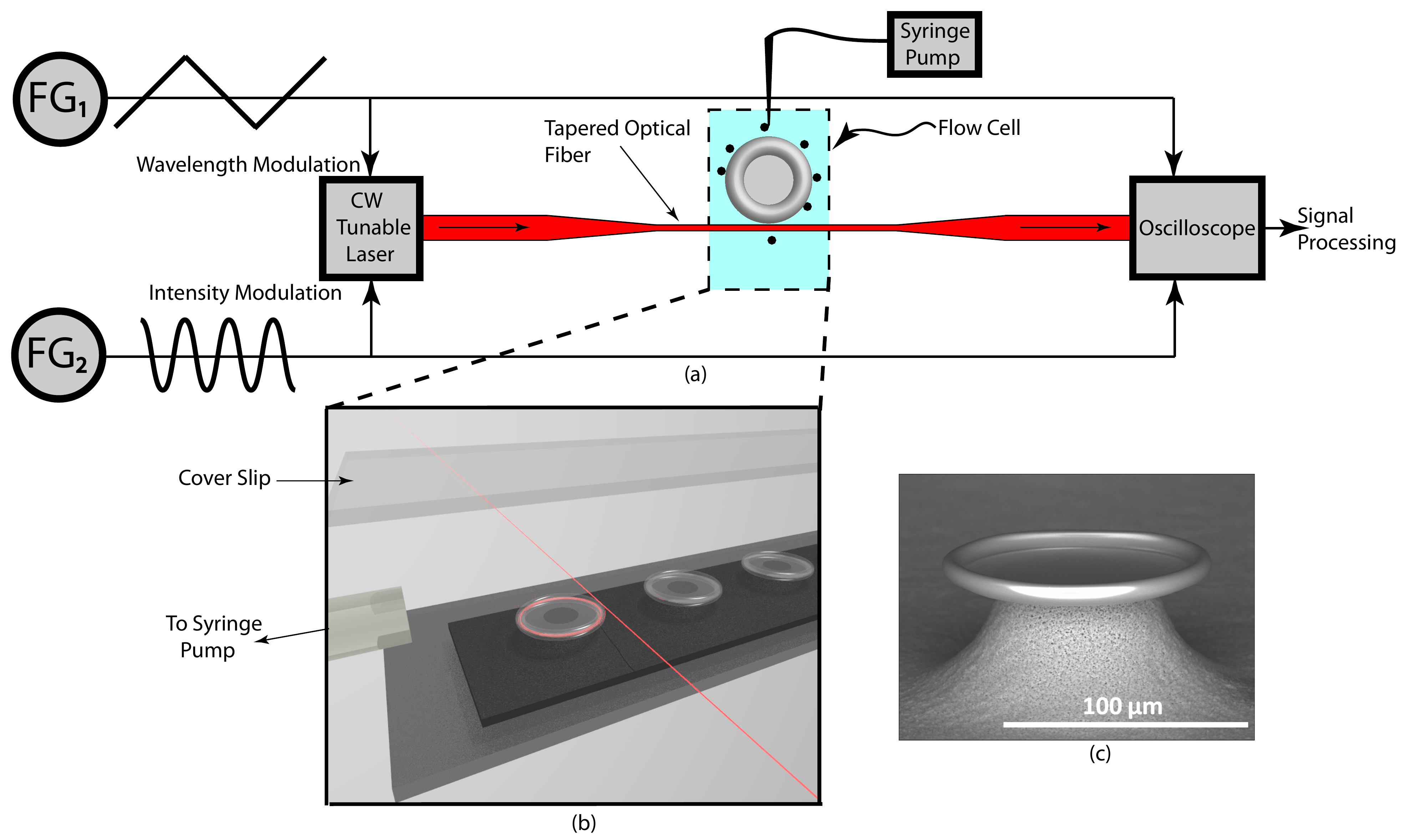}
\caption{(a) Experimental Setup, FG - Function Generator (b) Magnified view of the microacquarium. A cover slip is placed $2mm$ (approx.) above the  microcavity chip holder with a spacer to form the microacquarium. (c) Scanning electron microscope (SEM) image of the microtoroidal cavity }
\label{fig:setup}
\end{figure}

In the conventional PS-CRDS experiments with the free space cavities, a locking technique, such as Pound-Drever-Hall \cite{Drever_1983} technique, is used for locking the laser frequency to the cavity mode. Such locking techniques can also be applied to microcavities \cite{Schliesser_2010}. However,  locking the laser to the microcavity for biosensing means that information about the shift in resonant wavelength will not be measured;  at the same time, if the locking technique is not used, then the ring down time can not be measured because the microcavity resonance wavelength will shift due to the {biodetection} event. We have developed a novel experimental setup to overcome these difficulties. The experimental setup is shown in Fig.  \ref{fig:setup}.

\begin{figure}[htbp]
\centering
\subfigure[Frequency domain measurements ]
{
    \includegraphics[width=6.3cm]{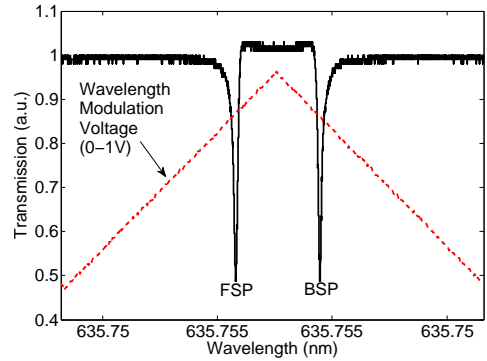}
}
\hspace{.1cm}
\subfigure[Time domain measurements ]
{
    \includegraphics[width=6.3cm]{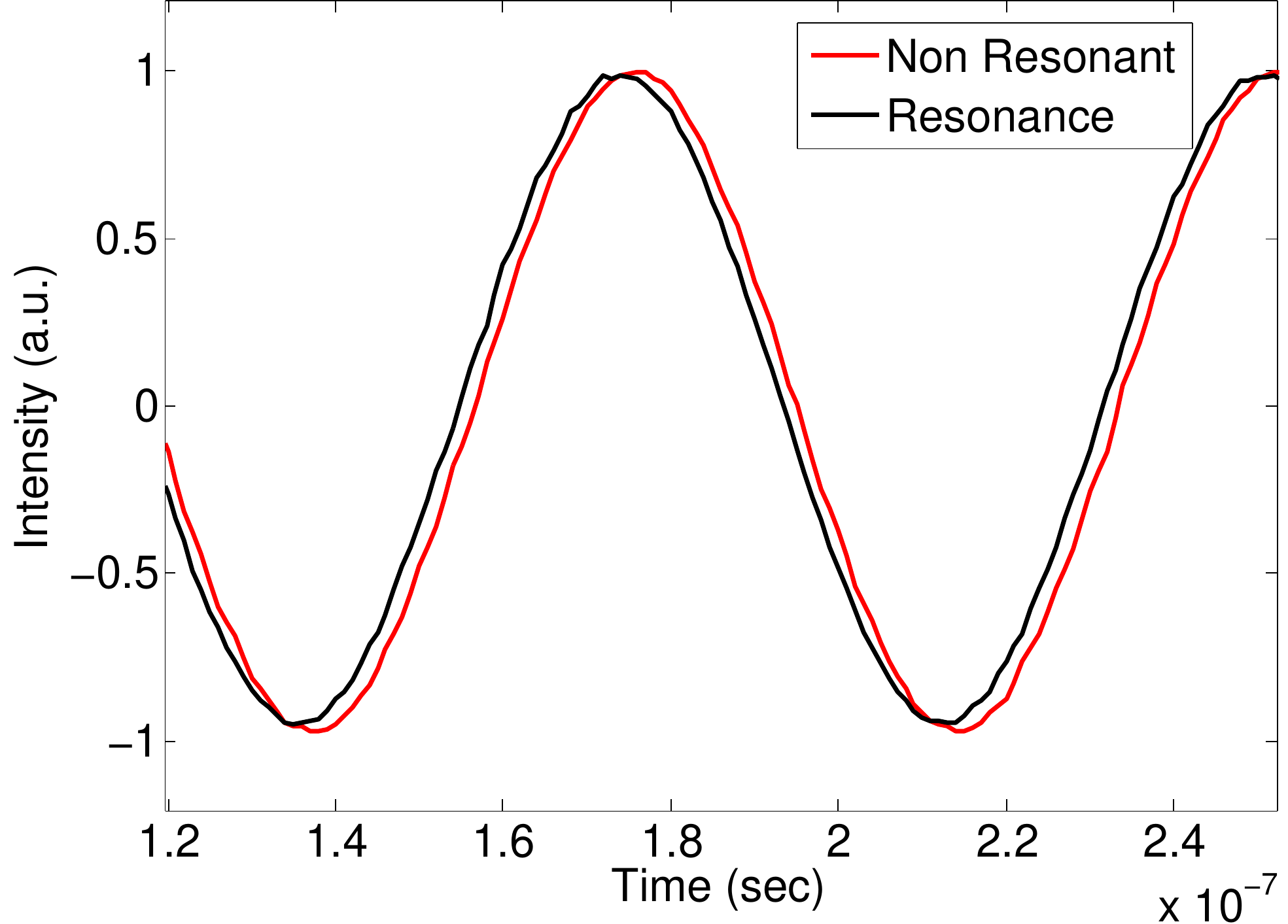}
}
\caption{Two types of measurements (a) Minimum of one of the resonant peak is tracked as function of the {biodetection} event (b) Phase shift experienced by the sinusoid coming out of the microcavity is tracked as a function of the {biodetection} event. With reference to Fig. (a), black sinusoid is extracted at the resonant peak.}
\label{fig:two_measurements}
\end{figure}

\begin{figure}[htbp]
\centering
\subfigure[Full Scan ]
{
    \includegraphics[width=6.3cm]{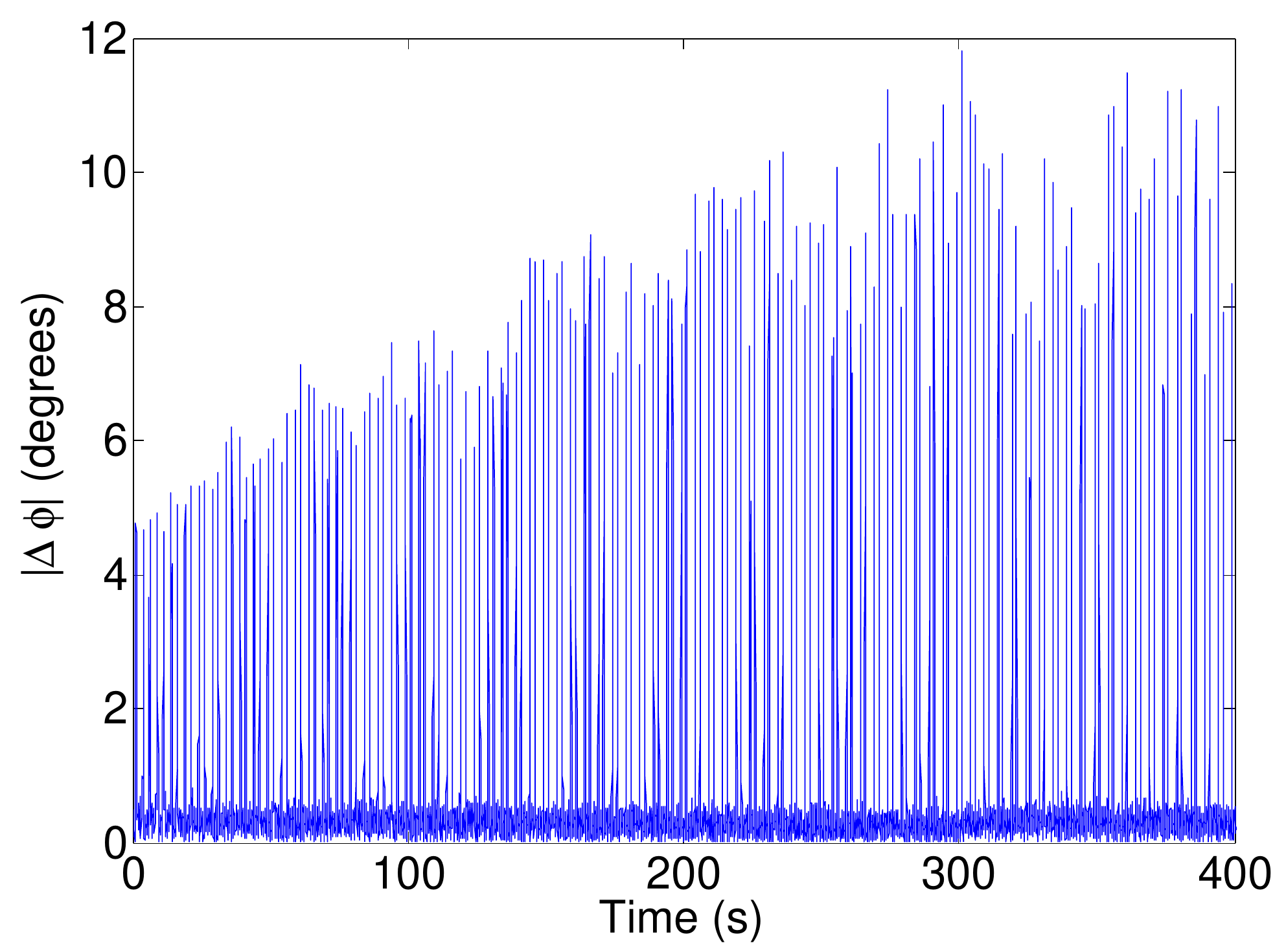}
}
\hspace{.1cm}
\subfigure[Zoom in view ]
{
    \includegraphics[width=5.8cm]{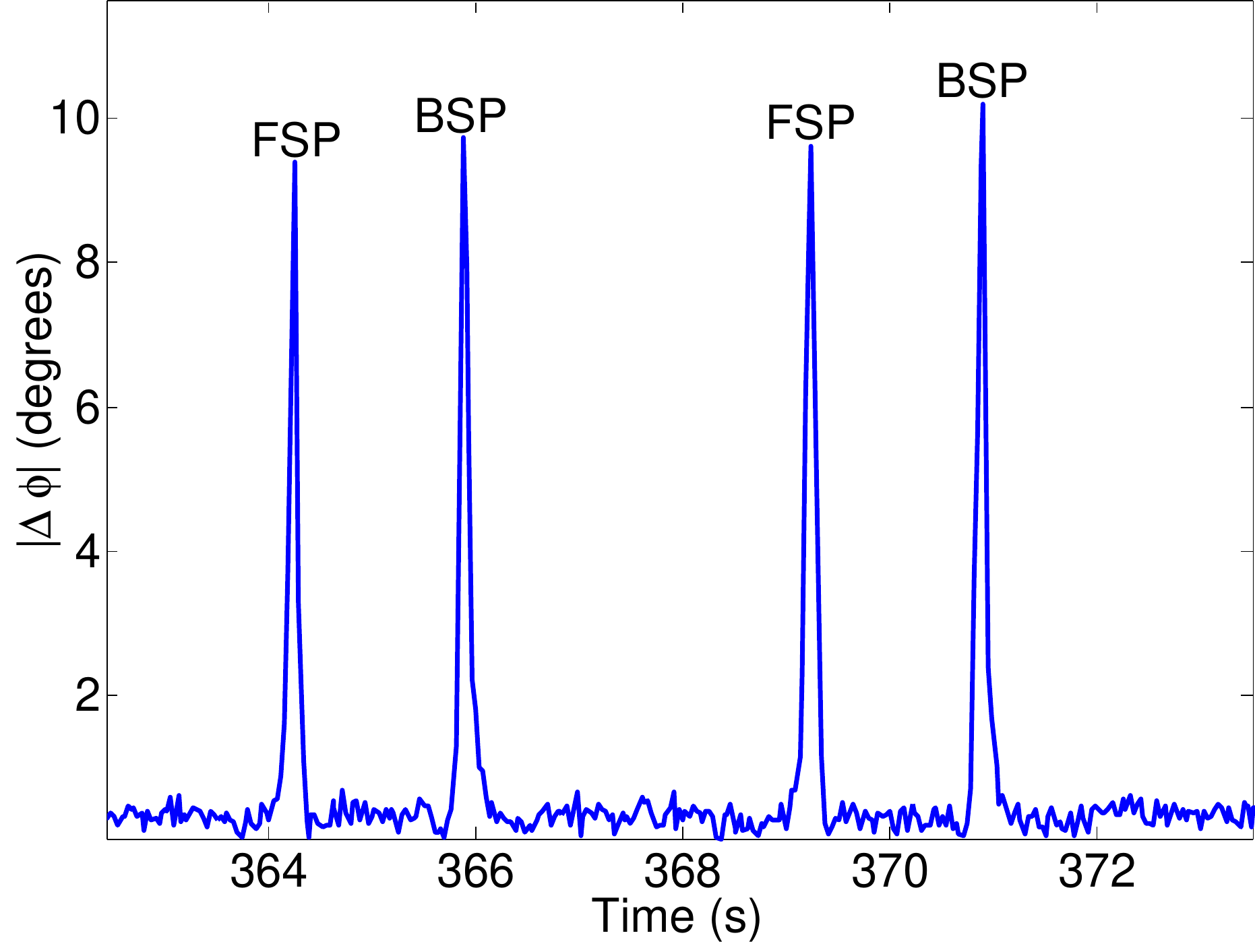}
}
\caption{Full scan with zoom in view of FSP and BSP (a) Continuous monitoring of phase as function of disassociation of the streptavidin  from the biotin (b) Due to limitations of the computer speed, two FSPs and BSPs are separated by 5s respectively.}
\label{fig:scan}
\end{figure}

A tapered fiber \cite{Knight_1997} is used to evanescently couple the light from the CW tunable laser centered at 633nm into the microtoroidal cavity with a $105\mu m$ ($6\mu m$)   major (minor) diameter. The microtoroidal cavity is covalently functionlized with biotin \cite{Hunt_2010}. The bioconjugated cavity is immersed in the PBS microacquarium to ensure the biostability and bioactivity of the biotin layer and subsequent protein,  and the setup is then flushed with 6mL of fresh PBS. We then inject $1nM$ of streptavidin via syringe pump at the rate of $50  \mu L/min$. After 7 minutes of the injection, we switch off the syringe pump and start tracking the change in the Q and the resonant wavelength. As shown in previous works \cite{Spillane_2004}, the quality factor is very dependent on the coupled input power, which is controlled by the coupling gap.  Therefore, all measurements are performed in contact at the same position on the cavity. We also monitor depth of the peaks to make sure that it remains constant throughout the recording phase. Therefore, changes in coupled input power do not play a role in the change in the Q \cite{Spillane_2004} in the present measurements. The procedure of tracking the ring down time and the resonant wavelength involves the following process:

A function generator ($FG_1$) with triangular wave output (100mHz, 1V peak-to-peak) continuously modulates the wavelength of a tunable laser (New Focus, Velocity Scan 633nm tunable laser) to generate the resonant peaks as shown in Fig. \ref{fig:two_measurements}a.
The peaks are labelled as Forward Scanning Peaks (FSP), and Backward Scanning Peaks (BSP). After locating a resonant peak, a second function generator ($FG_2$) is switched on (13MHz, 4V peak-to-peak) to provide sinusoidal modulation of the laser intensity.  After switching on both of the function generators, the phase shift between a reference sinusoid (sinusoidal modulated intensity of the laser before coupling) and the sinusoid at the fiber output is continuously recorded.  In such a scan, there will be phase shifts only at the resonant peaks (Fig. \ref{fig:scan}).
The same values of the phase shifts at the FSP and BSP indicate that there is no difference in the FWHM widths of the two peaks and hence any nonlinear effects are negligible\cite{Ilchenko_1992}, an important consideration for the biosensing experiments. The values of the peaks are the phase shifts which correspond to the ring down time (and hence Q) and positions of the peaks correspond to the resonant wavelength. Both of them can easily be extracted by further signal processing of the captured data.
\section{Results and discussion}
\begin{figure}[htbp]
\centering
\subfigure[Quality factor ]
{
    \includegraphics[width=6.3cm]{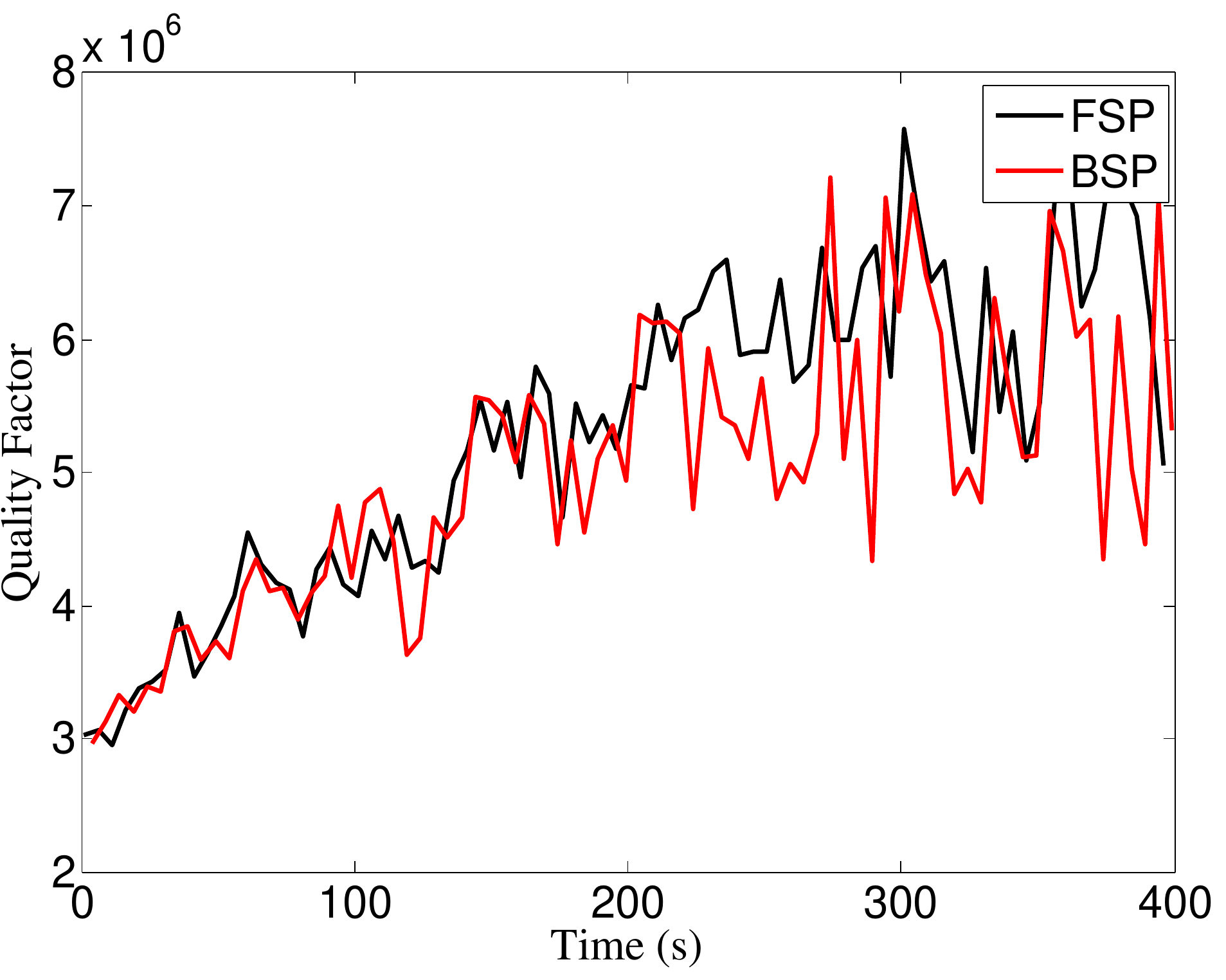}
}
\hspace{.1cm}
\subfigure[{Wavelength shift} ]
{
    \includegraphics[width=6.3cm]{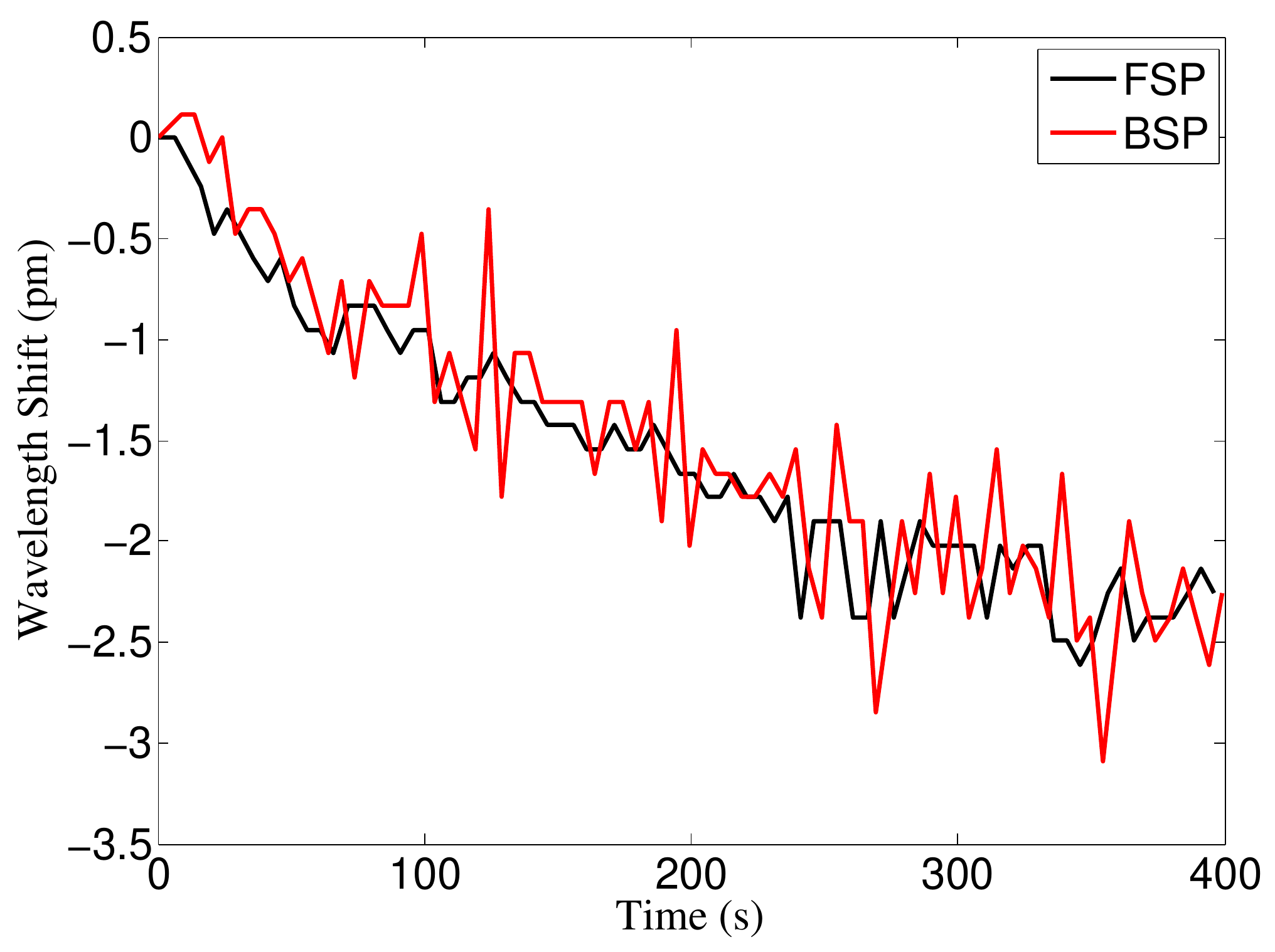}
}
\caption{Experimental parameters. The curves are not smooth because each data point is 5s apart on each curve, a limitation of the computer used in the experiment. In (b), negative values signify that the resonant wavelength is shifting towards blue}
\label{fig:para}
\end{figure}

\begin{figure}[htbp]
\centering\includegraphics[width=10cm]{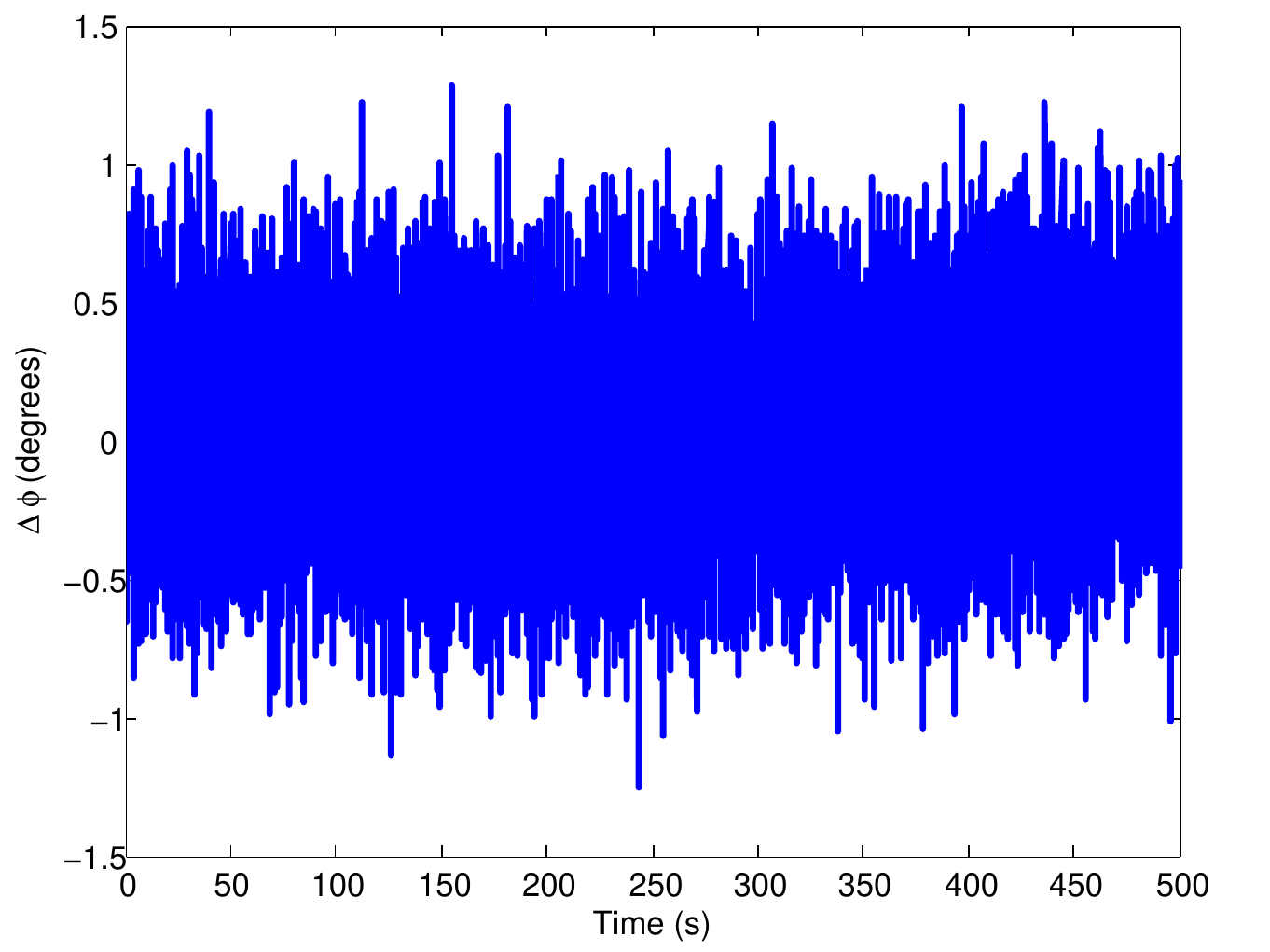}
\caption{A typical error signal, recorded by moving the taper far away from the microcavity. Mean: $\pm 0.2662^o$, Variance: $0.1076^o$, mode: $\pm 0.3387^o$. Based upon the detected signals,  $\left(\frac{Signal}{Noise}\right)_{min}=12$dB , and $\left(\frac{Signal}{Noise}\right)_{max}= 15.8$dB }
\label{fig:error}
\end{figure}

The binding kinetics of the biotin-streptavidin system in association and disassociation phase can be approximated by an exponential curve for the resonant wavelength \cite{Zhao_1992}. During the association phase, the resonant wavelength shifts towards the red and in the disassociation phase it shifts towards the blue \cite{Soteropulos_2011}. The change in resonant wavelength during the disassociation of the streptavidin from the biotin are shown in Fig. \ref{fig:para} which is in excellent agreement with \cite{Soteropulos_2011}. The change in Q during the disassociation phase is also shown which is following an increasing trend which makes sense as the binding and unbinding of the streptavidin will decrease and increase the overall quality factor of the microtoroid respectively.
The error signal, which is recorded with the tapered optical fiber alone, is also shown in Fig.  \ref{fig:error}. The noise shown in Fig. \ref{fig:error} is significantly smaller than the signal shown in Fig. \ref{fig:scan} with the minimum and maximum signal to noise ratios of 12dB and 15.8dB respectively.

The results indicate successful proof of the concept for the application of PS-CRDS to microcavities for biosensing. The PS-CRDS biosensor simultaneously tracks the Q and the resonant wavelength as a function of the {biodetection} event as opposed to its frequency domain counterpart which can only track the resonant wavelength. In our current setup, only tracking of the Q is free from noise due to the intensity fluctuations. However, the performance of the sensor can be improved by employing a lock-in amplifier which will reduce the error signal to $0.01^o$ and hence the minimum and maximum signal to noise ratios will be improved to $27$dB and $30$dB respectively. In conventional wavelength shift based microcavity biosensors, minimum of the resonant wavelength is tracked by software. On the other hand, by use of lock-in amplifier in PS-CRDS biosensor, accuracy of tracking of the resonant wavelength will be improved as compared to the frequency domain biosensor.

Since absorption of analytes influences the quality factor of the microtoroid, the PS-CRDS biosensor can also conduct such measurements in a faster (electronics) and an intensity noise free manner. On the other hand, absorption measurements \cite{Armani_2006,Rosenberger_2007} in the wavelength shift based microcavity sensors are slow (software) and carry noise not only due to the intensity fluctuations but also due to fitting of Lorentzian curves to the resonant peaks. Moreover, other noise mechanisms that are present in the wavelength shift based microcavity sensors, such as laser frequency jitter, conversion of $FG_1$ voltage to the wavelength \cite{Lu_2011}, and shift in the resonant wavelength due to thermal fluctuations \cite{Swaim_2011} are also minimized in the Q measurements. This observation is clear from the fact that these noise mechanisms will effect the location of the resonant peak whereas the  {Q measurements are independent of the peak's position (see Fig. 2b along with its caption, Eq. (\ref{Eqn:PS_CRDS}), and Eq. (\ref{Eqn:Q}))}.

CRDS with free space cavities is an emerging technique for liquid phase applications such as liquid chromatography \cite{Snyder_2003,Bahnev_2005}. However, such devices suffer from scattering and reflection losses due to the placement of the liquid sample inside the cavity. In some cases \cite{Bahnev_2005}, liquid is in contact with the mirrors which degrades the mirrors over a period of time. Moreover, different mirrors need to be employed to meet wavelength requirements in an experiment. The PS-CRDS sensor demonstrated in this work can overcome the above mentioned difficulties and has potential to offer a compact and sensitive solution for the liquid chromatography applications.
\section{Conclusions}
In summary, we have combined the advantages of CRDS (high noise immunity) and microcavities (high Q, low cost and compact) to develop a sensor that can simultaneously track the Q and the resonant wavelength as a function of a {biodetection} event. The sensor can also provide {faster} results for absorption measurements than its frequency domain counterpart. This work is also the first demonstration of applying PS-CRDS to microcavities for liquid phase applications. This PS-CRDS microcavity sensor has potential of finding wide applications not only in biosensing but also in analytical chemistry, pharmaceutical and agriculture sectors.

\section*{Acknowledgments}
This work was supported by the NSERC-CREATE training program in integrated sensor systems, McGill institute of advanced materials, Montreal, Canada and the National Science Foundation [085281]. The authors would also like to thank Raja Abdullah and Jason Gamba for useful discussions.
\bibliographystyle{osajnl}
\bibliography{references}

\begin{thebibliography}{10}
\newcommand{\enquote}[1]{``#1''}

\bibitem{Fan_2008}
X.~Fan, I.~M. White, S.~I. Shopoua, H.~Zhu, J.~D. Suter, and Y.~Sun,
  \enquote{{Sensitive optical biosensors for unlabeled targets: A review},}
  {Analytica Chimica Acta} \textbf{{620}}, {8--26} ({2008}).

\bibitem{Vollmer_2002}
F.~Vollmer, D.~Braun, A.~Libchaber, M.~Khoshsima, I.~Teraoka, and S.~Arnold,
  \enquote{{Protein detection by optical shift of a resonant microcavity},}
  {Applied Physics Letters} \textbf{{80}}, {4057--4059} ({2002}).

\bibitem{Washburn_2011}
A.~L. {Washburn} and R.~C. {Bailey}, \enquote{{Photonics-on-a-chip: recent
  advances in integrated waveguides as enabling detection elements for
  real-world, lab-on-a-chip biosensing applications},} The Analyst
  \textbf{136}, 227 (2011).

\bibitem{Soteropulos_2011}
C.~E. Soteropulos, H.~K. Hunt, and A.~M. Armani, \enquote{Determination of
  binding kinetics using whispering gallery mode microcavities,} Applied
  Physics Letters \textbf{99}, 103703 (2011).

\bibitem{Cheema_2011}
M.~I. Cheema and A.~G. Kirk, \enquote{{Application of ring down measurement
  approach to micro-cavities for bio-sensing applications},} in \enquote{{Proc.
  SPIE 7888, 788808 },} , {}, ed., {} ({}, {2011}), p.~{}. {}.

\bibitem{Ilchenko_2002}
V.~S. Ilchenko and L.~Maleki, \enquote{{High-Q whispering-gallery mode sensor
  in liquids },} in \enquote{{Proc. SPIE:Laser Resonators and Beam Control V
  4629, 72},} {} ({}, {2002}). {}.

\bibitem{Armani_2006}
A.~M. Armani and K.~J. Vahala, \enquote{Heavy water detection using
  ultra-high-q microcavities,} Opt. Lett. \textbf{31}, 1896--1898 (2006).

\bibitem{White_2008}
I.~M. White and X.~Fan, \enquote{On the performance quantification of resonant
  refractive index sensors,} Opt. Express \textbf{16}, 1020--1028 (2008).

\bibitem{Hu_2009}
J.~Hu, X.~Sun, A.~Agarwal, and L.~C. Kimerling, \enquote{Design guidelines for
  optical resonator biochemical sensors,} J. Opt. Soc. Am. B \textbf{26},
  1032--1041 (2009).

\bibitem{Anderson_1984}
D.~Anderson, J.~Frisch, and C.~Masser, \enquote{{Mirror reflectometer based on
  optical cavity decay time},} {Applied Optics} \textbf{{23}}, {1238--1245}
  ({1984}).

\bibitem{Keefe_1988}
A.~Okeefe and D.~Deacon, \enquote{{Cavity ring-down optical spectrometer for
  absorption-measurements using pulsed laser sources},} {Review of Scientific
  Instruments} \textbf{{59}}, {2544--2551} ({1988}).

\bibitem{Berden_2009}
E.~G. Berden and R.~Engeln, \emph{Cavity Ring-Down Specroscopy: Techniques and
  Applications} (Wiley, 2009).

\bibitem{Lehmann_2009}
K.~K. Lehmann and H.~Huang, \emph{Frontiers of Molecular Spectroscopy}
  (Elsevier, 2009), chap. Optimal Signal Processing in Cavity Ring-Down
  Spectroscopy, pp. 623--658.

\bibitem{Herbelin_1980}
J.~M. Herbelin, J.~A. McKay, M.~A. Kwok, R.~H. Ueunten, D.~S. Urevig, D.~J.
  Spencer, and D.~J. Benard, \enquote{Sensitive measurement of photon lifetime
  and true reflectances in an optical cavity by a phase-shift method,} Appl.
  Opt. \textbf{19}, 144--147 (1980).

\bibitem{Engeln_1996}
R.~Engeln, G.~VonHelden, G.~Berden, and G.~Meijer, \enquote{{Phase shift cavity
  ring down absorption spectroscopy},} {Chemical Physics Letters}
  \textbf{{262}}, {105--109} ({1996}).

\bibitem{Sneppen_2008}
L.~van~der Sneppen, \enquote{Liquid-phase cavity ring-down spectroscopy for
  improved analytical detection sensitivity,} Ph.D. thesis, Vrije University,
  Amsterdam (2008).

\bibitem{Tarsa_2008}
T.~Peter and L.~Kevin, \emph{Optical Biosensors: Today and Tomorrow} (Elsevier
  B.V, 2008), chap. Cavity ring Down biosensing, pp. 403--418.

\bibitem{Tarsa_2004}
P.~Tarsa, A.~Wist, P.~Rabinowitz, and K.~Lehmann, \enquote{{Single-cell
  detection by cavity ring-down spectroscopy},} {Applied Physics Letters}
  \textbf{{85}}, {4523--4525} ({2004}).

\bibitem{Loock_2004}
Z.~Tong, A.~Wright, T.~McCormick, R.~Li, R.~D. Oleschuk, and H.-P. Loock,
  \enquote{Phase-shift fiber-loop ring-down spectroscopy,} Analytical Chemistry
  \textbf{76}, 6594--6599 (2004). PMID: 15538782.

\bibitem{Barnes_2008}
J.~Barnes, B.~Carver, J.~M. Fraser, G.~Gagliardi, H.~P. Loock, Z.~Tian,
  M.~W.~B. Wilson, S.~Yam, and O.~Yastrubshak, \enquote{{Loss determination in
  microsphere resonators by phase-shift cavity ring-down measurements},}
  {Optics Express} \textbf{{16}}, {13158--13167} ({2008}).

\bibitem{Hunt_2010}
H.~K. Hunt, C.~Soteropulos, and A.~M. Armani, \enquote{Bioconjugation
  strategies for microtoroidal optical resonators,} Sensors \textbf{10},
  9317--9336 (2010).

\bibitem{Armani_2003}
D.~K. Armani, T.~J. Kippenberg, S.~M. Spillane, and K.~J. Vahala,
  \enquote{Ultra-high-{Q} toroid microcavity on a chip,} Nature \textbf{421}
  (2003).

\bibitem{Spillane_2004}
S.~Spillane, \enquote{Fiber-coupled ultra-high-q microresonators for nonlinear
  and quantum optics,} Ph.D. thesis, CALTECH (2004).

\bibitem{Drever_1983}
R.~W.~P. Drever, J.~L. Hall, F.~V. Kowalski, J.~Hough, G.~M. Ford, A.~J.
  Munley, and H.~Ward, \enquote{Laser phase and frequency stabilization using
  an optical resonator,} Applied Physics B Photophysics And Laser Chemistry
  \textbf{31}, 97--105 (1983).

\bibitem{Schliesser_2010}
A.~Schliesser and T.~J. Kippenberg, \emph{Advances in Atomic Molecular and
  Optical Physics} (Elsevier Inc., 2010), vol.~58, chap. Chapter 5 – Cavity
  Optomechanics with Whispering-Gallery Mode Optical Micro-Resonators, pp.
  207--323.

\bibitem{Knight_1997}
J.~C. Knight, G.~Cheung, F.~Jacques, and T.~A. Birks, \enquote{Phase-matched
  excitation of whispering-gallery-mode resonances by a fiber taper,} Opt.
  Lett. \textbf{22}, 1129--1131 (1997).

\bibitem{Ilchenko_1992}
V.~S. Ilchenko and M.~L. Gorodetskii, \enquote{Thermal nonlinear effects in
  optical whispering gallery microresonators,} Laser Physics \textbf{2},
  1004--1009 (1992).

\bibitem{Zhao_1992}
S.~Zhao and W.~M. Reichert, \enquote{Influence of biotin lipid surface density
  and accessibility on avidin binding to the tip of an optical fiber sensor,}
  Langmuir \textbf{8}, 2785--2791 (1992).

\bibitem{Rosenberger_2007}
A.~T. Rosenberger, \enquote{Analysis of whispering-gallery microcavity-enhanced
  chemical absorption sensors,} Opt. Express \textbf{15}, 12959--12964 (2007).

\bibitem{Lu_2011}
T.~Lu, H.~Lee, T.~Chen, S.~Herchak, J.-H. Kim, S.~E. Fraser, R.~C. Flagan, and
  K.~Vahala, \enquote{High sensitivity nanoparticle detection using optical
  microcavities,} Proceedings of the National Academy of Sciences  (2011).

\bibitem{Swaim_2011}
J.~D. Swaim, J.~Knittel, and W.~P. Bowen, \enquote{Detection limits in
  whispering gallery biosensors with plasmonic enhancement,} Applied Physics
  Letters \textbf{99}, 243109 (2011).

\bibitem{Snyder_2003}
K.~L. Snyder and R.~N. Zare, \enquote{Cavity ring-down spectroscopy as a
  detector for liquid chromatography,} Analytical Chemistry \textbf{75},
  3086--3091 (2003).

\bibitem{Bahnev_2005}
B.~Bahnev, L.~van~der Sneppen, A.~E. Wiskerke, F.~Ariese, C.~Gooijer, and
  W.~Ubachs, \enquote{Miniaturized cavity ring-down detection in a liquid flow
  cell,} Analytical Chemistry \textbf{77}, 1188--1191 (2005).

\end{thebibliography}
\end{document}